\documentclass[aps,prl,twocolumn,groupedaddress,amsmath,amssymb]{revtex4}

\usepackage{graphicx}% Include figure files
\usepackage{dcolumn}% Align table columns on decimal point
\usepackage{bm}% bold math

\begin{document}

\title{Doping-induced superconductivity in the topological semimetal Mo$_{5}$Si$_{3}$}

\date{\today}

\author{Jifeng Wu$^{1,2,3}$\footnote{These authors contributed equally to this work.}}
\author{Chenqiang Hua$^{4*}$}
\author{Bin Liu$^{1,2,3}$}
\author{Yanwei Cui$^{1,2,4}$}
\author{Qinqing Zhu$^{1,2,3}$}
\author{Guorui Xiao$^{1,2,4}$}
\author{Siqi Wu$^{4}$}
\author{Guanghan Cao$^{4}$}
\author{Yunhao Lu$^{4,5}$}
\author{Zhi Ren$^{1,2}$\footnote{renzhi@westlake.edu.cn}}

\affiliation{$^{1}$School of Science, Westlake University, 18 Shilongshan Road, Hangzhou 310064, P. R. China}
\affiliation{$^{2}$Institute of Natural Sciences, Westlake Institute for Advanced Study, 18 Shilongshan Road, Hangzhou 310064, P. R. China}
\affiliation{$^{3}$Department of Physics, Fudan University, Shanghai 200433, P. R. China}
\affiliation{$^{4}$Zhejiang Province Key Laboratory of Quantum Technology and Device, Department of Physics, Zhejiang University, Hangzhou 310027, P. R. China}
\affiliation{$^{5}$State Key Lab of Silicon Materials, School of Materials Science and Engineering, Zhejiang University, Hangzhou 310027, P. R. China}

\begin{abstract}
Chemical doping of topological materials may provide a possible route for realizing topological superconductivity. However, all such cases known so far are based on chalcogenides. Here we report the discovery of superconductivity induced by Re doping in the topological semimetal Mo$_{5}$Si$_{3}$ with a tetragonal structure.
Partial substitution of Re for Mo in Mo$_{5-x}$Re$_{x}$Si$_{3}$ results in an anisotropic shrinkage of the unit cell up to the solubility limit of approximately $x$ = 2.
Over a wide doping range (0.5 $\leq$ $x$ $\leq$ 2), these silicides are found to be weakly coupled superconductors with a fully isotropic gap. $T_{\rm c}$ increases monotonically with $x$ from 1.67 K to 5.78 K, the latter of which is the highest among superconductors of the same structural type. This trend in $T_{\rm c}$ correlates well with the variation of the number of valence electrons, and is mainly ascribed to the enhancement of electron-phonon coupling.
In addition, band structure calculations reveal that superconducting Mo$_{5-x}$Re$_{x}$Si$_{3}$ exhibits nontrivial band topology characterized by $Z_{2}$ invariants (1;000) or (1;111) depending on the Re doping level.
Our results suggest that transition metal silicides are a fertile ground for the exploration of candidate topological superconductors.
\end{abstract}

\maketitle
\maketitle

\section{I. Introduction}
In the past decade, topological superconductors (TSC) have received widespread attention as a new quantum state of matter \cite{TSCreview1,TSCreview2,TSCreview3}.
The most prominent character of TSC is the existence of Majorana fermions \cite{TSCreview1,TSCreview2}. These fermions are their own antiparticles and obey non-Abelian statistics, which makes TSC useful in fault-tolerant quantum computing.
In theory, in addition to topological materials that are intrinsically superconducting \cite{superconductingTM1,superconductingTM2,superconductingTM3,superconductingTM4}, doped topological materials that exhibit superconductivity are also natural candidates for TSC \cite{TSCreview3}.
In reality, examples in the latter category include $A_{x}$Bi$_{2}$Se$_{3}$ ($A$ = Cu, Sr, Nb) \cite{CuxBi2Se3,SrxBi2Se3,NbxBi2Se3}, Tl$_{0.6}$Bi$_{2}$Te$_{3}$ \cite{TlxBi2Te3}, Sn$_{1-x}$In$_{x}$Te \cite{Sn1-xInxTe}, K- and [C$_{6}$H$_{11}$N$_{2}$]-intercalated WTe$_{2}$ \cite{KxWTe2,organicintercalated}. These materials are all based on transition metal or main-group chalcogenides, and most of their parent compounds are either topological insulators or topological crystalline insulators.
In comparison, doping-induced superconductivity has been rarely observed in topological semimetals.
Very recently, CoSi was shown to have a topologically nontrivial band structure \cite{CoSi1,CoSi2}, suggesting that TSC may also be found in transition metal silicides.
However, no such example has been reported to date. Note that, compared with chalcogenides, the transition metal silicides generally have higher hardness and thermal stability \cite{highhardness}, and thus are better suited for practical applications.

Several theoretical calculations consistently show that Mo$_{5}$Si$_{3}$ is a high symmetry line topological semimetal \cite{Mo5Si3prediction1,Mo5Si3prediction2,Mo5Si3prediction3}.
This material belongs to the tetragonal $D$8$_{m}$ category of the $A_{5}$$B_{3}$ family, and its structure is sketched in Figs. 1(a) and (b).
In the tetragonal unit cell, there are two distinct Mo sites Mo(1) and Mo(2), the nearest-neighbours of which form zigzag and linear chains running along the $c$-axis, respectively \cite{Mo5Si3structure}.
Since the density of states at the Fermi level [$N$($E_{\rm F}$)] of Mo$_{5}$Si$_{3}$ is dominated by the Mo 4$d$ states, this one-dimensional arrangement of the Mo atoms results in significant anisotropy in physical properties including thermal expansion and resistivity \cite{Mo5Si3anisotropy}. However, Mo$_{5}$Si$_{3}$ has long been studied as a high-temperature structural alloy \cite{HTalloy1,HTalloy2,HTalloy3}, while little attention has been paid to its low temperature properties.
In this respect, it is worth noting that a number of compounds isostructural to Mo$_{5}$Si$_{3}$ are found superconducting and their $T_{\rm c}$ correlates with the average number of valence electrons per atom ratio ($e$/$a$) \cite{Nb5Sn2Al}. The $e$/$a$ value falls in the range between 4.25 and 4.625 for most of these superconductors, but is significant higher (5.25) for W$_{5}$Si$_{3}$ that is isoelectronic to Mo$_{5}$Si$_{3}$.
This suggests the presence of another superconducting region at higher $e$/$a$ values, which is also expected by the Matthias rule \cite{MTrule}.
It is thus speculated that superconductivity may be induced in Mo$_{5}$Si$_{3}$ by increasing the number of valence electrons.
\begin{figure*}
\includegraphics*[width=17cm]{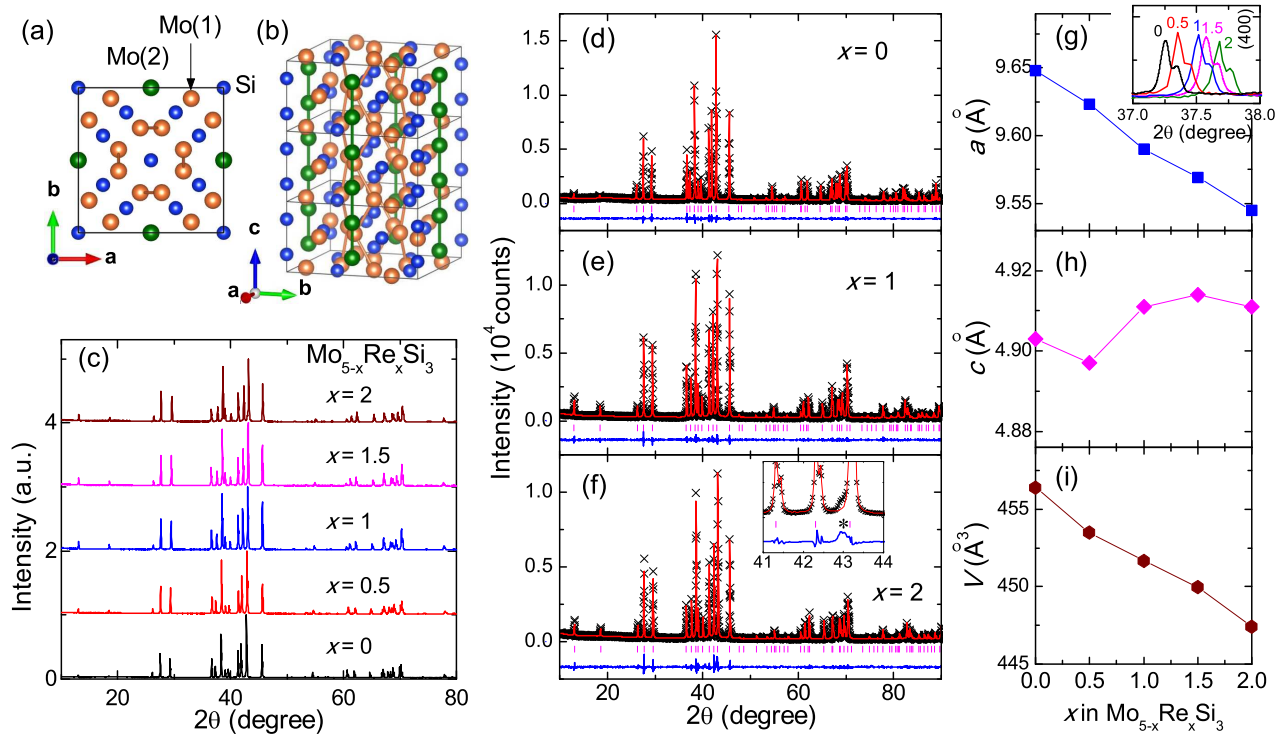}
\caption{
(a) Schematic structure of Mo$_{5}$Si$_{3}$ projected perpendicular to the $c$-axis.
(b) A three-dimensional view of the Mo$_{5}$Si$_{3}$ unit cells.
(c) Powder x-ray diffraction patterns for the series of Mo$_{5-x}$Re$_{x}$Si$_{3}$ samples.
(d)-(f) Structural refinement results for $x$ = 0, 1 and 2, respectively. The inset in panel (f) shows a zoom of the data for $x$ = 2 in the 2$\theta$ range between 41$^{\circ}$ and 44$^{\circ}$, and the impurity peak near 2$\theta$ $\approx$ 43$^{\circ}$ is marked by the asterisk.
(g-i) Re content $x$ dependence of the lattice constants and unit-cell volume for the Mo$_{5-x}$Re$_{x}$Si$_{3}$ samples, respectively.
The inset in (g) shows the (400) peak for these samples.
}
\label{fig1}
\end{figure*}

With this in mind, we employ Re as a dopant since it has one more valence electron than Mo.
Structural analysis indicates that Mo$_{5-x}$Re$_{x}$Si$_{3}$ retains a single tetragonal structure up to $x$ $\approx$ 2, and the increase of Re content $x$ results in a reduction of the $a$-axis but has little affect on the $c$-axis. For $x$ $\geq$ 0.5, bulk superconductivity is observed with a maximal $T_{\rm c}$ of 5.78 K observed at the Re solubility limit.
The dependencies of $T_{\rm c}$ on $N$($E_{\rm F}$), electron-phonon coupling constant as well as $e$/$a$ are examined, and various superconducting parameters are obtained.
Moreover, the evolution of band topology with Re doping is investigated by theoretical calculations.
The implication of these results on the possibility of topological superconductivity in Re-doped Mo$_{5}$Si$_{3}$ is discussed.

\section{II. Materials and Methods}
\begin{table}[b]
\caption{Atomic coordinates for Mo$_{5-x}$Re$_{x}$Si$_{3}$.}
\renewcommand\arraystretch{1.3}
\begin{tabular}{p{1.5cm}<{\centering}p{1.5cm}<{\centering}p{0.9cm}<{\centering}p{0.9cm}<{\centering}p{0.9cm}<{\centering}p{1.9cm}<{\centering}}
\\
\hline % Top horizontal line
   Atoms & site  &  $x$  & $y$ & $z$ & Occupancy \\

\hline % In-table horizontal line
Mo(1)/Re(1)							& 	 16$k$ 	 & 0.074	& 0.223 & 0 & (1-0.2$x$)/0.2$x$ 	\\
Mo(2)/Re(2)							& 	  4$b$ 	 & 0 & 0.5 & 0.25 & (1-0.2$x$)/0.2$x$		\\
Si(1)				&      8$h$    & 			0.17 & 0.67 & 0 & 1 \\
Si(2)	&      4$a$ &   0  & 0 & 0.25 & 1  \\
\hline% In-table horizontal line
\hline % Bottom horizontal line
\end{tabular}
\label{Table3}
\end{table}
Polycrystalline Mo$_{5-x}$Re$_{x}$Si$_{3}$ samples with $x$ = 0, 0.5, 1, 1.5 and 2 were prepared by the arc melting method.
High purity powders of Mo (99.9\%), Re (99.9\%) and Si (99.99\%) were weighed according to the stoichiometric ratio, mixed thoroughly using a mortar and pestle, and pressed into pellets in an argon-filled glovebox. The pellets were melted several times in an arc furnace under argon atmosphere, followed by rapid cooling on a water-chilled copper plate.
The resulting ingots were grounded, pressed gain into pellets, and annealed at 1600 $^{\circ}$C in a vacuum furnace for 24 hours, followed by furnace cooling to room temperature.
We have examined carefully ingots and pellets of all samples but no single crystals could be obtained.
\begin{figure*}
\includegraphics*[width=17.5cm]{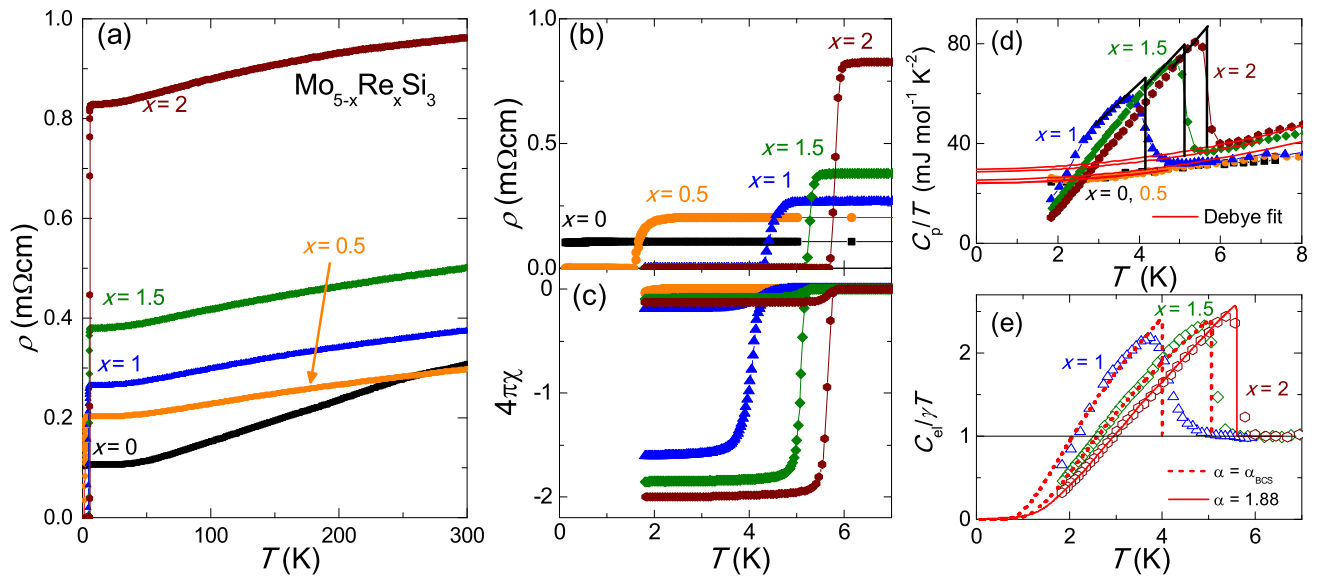}
\caption{
(a) Temperature dependence of resistivity for the series of Mo$_{5-x}$Re$_{x}$Si$_{3}$ samples.
(b) A zoom of the same set of data in the low temperature region.
(c) Magnetic susceptibility data in the same temperature range as in (b) for these samples.
(d) Low temperature $C_{\rm p}$ data for the series of Mo$_{5-x}$Re$_{x}$Si$_{3}$ samples.
The solid lines are entropy conserving constructions to estimate the specific heat jump as well as $T_{\rm c}$ and dashed lines are fits by the Debye model.
(e) Normalized electronic specific heat for the samples with $x$ = 1, 1.5 and 2. The dashed and solid lines are fits by the $\alpha$ model.
}
\label{fig1}
\end{figure*}

The phase purity was checked by powder x-ray diffraction (XRD) at room temperature using a Bruker D8 Advance diffractometer with Cu K$\alpha$ radiation. The chemical composition was measured with an energy-dispersive x-ray (EDX) spectrometer affiliated to a Hitachi field emission scanning electron microscope (SEM). Electrical resistivity and specific heat measurements were carried out in a Quantum Design Physical Property Measurement System (PPMS-9 Dynacool). The resistivity was measured using a standard four-probe method down to 1.8 K, and, if necessary, then from 1.8 to 150 mK with an adiabatic dilution refrigerator option. The applied current is 1 mA. The specific heat was measured down to 1.8 K using a relaxation method. Magnetic susceptibility measurements down to 1.8 K were performed in a Quantum Design Magnetic Property Measurement System (MPMS3) with an applied magnetic field of 0.5 mT in both zero-field cooling (ZFC) and field cooling (FC) modes. For consistency, all the results were collected on the same sample for each $x$ value.

First-principles band structure calculations were performed using the Perdew-Burke-Ernzerhof (PBE)\cite{PBE} exchange-correlation functional in the Vienna ab-initio simulation package \cite{VASP}. Our convergence threshold of Hellmann-Feynman force is 0.01 eV/{\AA} and the energy convergence criteria was set to 10$^{-6}$ eV. In both optimization and self-consistent calculations, the energy cutoff was fixed to 450 eV. Besides, we adopted 9$\times$9$\times$9 $k$
points to mesh the Brillouin zone. The effect of spin-orbital coupling (SOC) \cite{SOC} is included for Wannier-function-based calculations of $Z_{2}$ invariants \cite{SOC,wanier} .
\begin{figure*}
\includegraphics*[width=17.5cm]{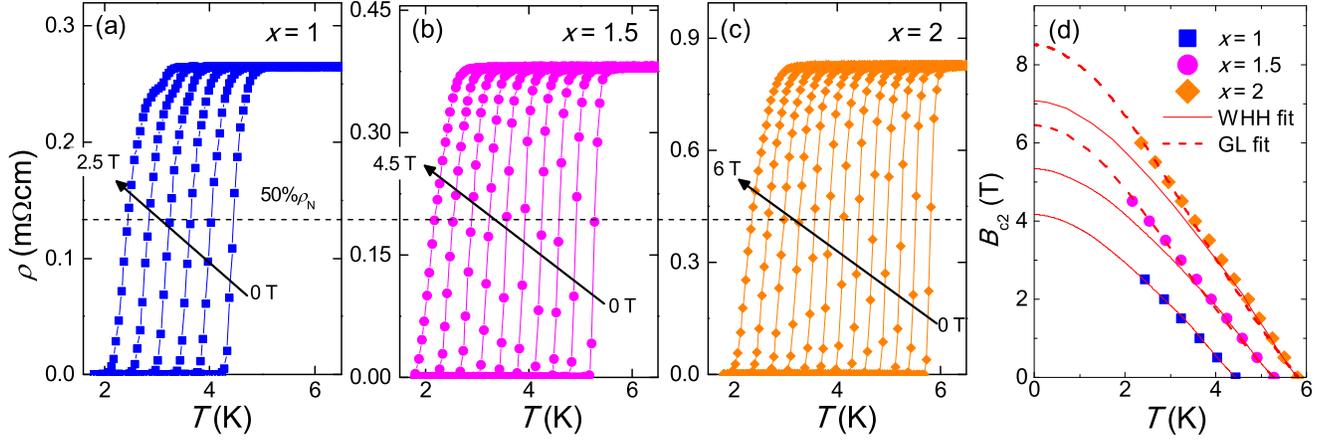}
\caption{
(a-c) Temperature dependence of resistivity under various fields for the Mo$_{5-x}$Re$_{x}$Si$_{3}$ samples with $x$ = 1, 1.5 and 2, respectively.
In each panel, the field increment is 0.5 T.
(d) Temperature dependence of upper critical field for these samples. The solid and dashed lines are fits by the WHH and Ginzburg-Landau models, respectively.
}
\label{fig1}
\end{figure*}

\section{III. Results and Discussion}
Fig. 1(c) shows the results of XRD measurements for the series of Mo$_{5-x}$Re$_{x}$Si$_{3}$ samples.
It is obvious that all the patterns are very similar, indicating that the structure remains unchanged upon Re doping.
For the refinement, the $I$4/$mcm$ space group is employed and the Re atoms are assumed to occupy disorderly the Mo(1) and Mo(2) sites (see Table I) \cite{orderedmodel}.
As exemplified in Figs. 1(d)-(f), a good agreement is obtained between the calculated and observed patterns (the CIF files are available through the ICSD database, CSD 2027050-2027052) and the refinement results are listed in Table II.
For $x$ = 2, a small impurity peak is discernible at 2$\theta$ $\approx$ 43$^{\circ}$ [see the inset of Fig. 1(f)], suggesting that this $x$ value is very close to the solubility limit of Re in Mo$_{5}$Si$_{3}$.
Figures 2(d)-(f) show the refined lattice parameters and unit-cell volume ($V$) plotted as a function of Re content $x$.
The lattice parameters for Mo$_{5}$Si$_{3}$ ($x$ = 0) are $a$ = 9.648(1) {\AA} and $c$ = 4.903(1) {\AA}, in good agreement with the previous report \cite{Mo5Si3structure}.
With increasing $x$, the $a$-axis decreases by $\sim$1.1 \% to 9.545(1) {\AA}, which is clearly evidenced by the shift of (400) peak position toward higher 2$\theta$ values [see the inset of Fig. 1(g)].
In comparison, the variation of $c$-axis is within $\sim$0.2 \%.
As a result, $V$ shrinks by nearly 2\% at $x$ = 2 compared with $x$ = 0, which is as expected since Re has a smaller atomic radius than Mo \cite{radius}.

The morphology of these Mo$_{5-x}$Re$_{x}$Si$_{3}$ samples was also investigated by SEM.
At low $x$ values, the samples contains disconnected grains with a size of several tenth $\mu$m.
As the Re content increases, the grains become interconnected with the voids getting fewer and smaller.
This signifies a growth in the grain size, probably due to a reduction in the energy barrier caused by Re doping.
The EDX spectra were collected from several locations for each sample (see Fig. S1 of the Supporting Information).
Averaging the data yields the measured chemical compositions of Mo$_{5.06(8)}$Si$_{2.93(8)}$, Mo$_{4.54(11)}$Re$_{0.47(9)}$Si$_{2.99(9)}$, Mo$_{4.01(11)}$Re$_{1.09(3)}$Si$_{2.90(12)}$, Mo$_{3.65(6)}$Re$_{1.43(3)}$Si$_{2.91(7)}$, and Mo$_{3.17(13)}$Re$_{1.83(6)}$Si$_{2.99(11)}$ for $x$ = 0, 0.5, 1, 1.5 and 2, respectively, which agree with the nominal ones within the experimental error.

\begin{figure*}
\includegraphics*[width=17.5cm]{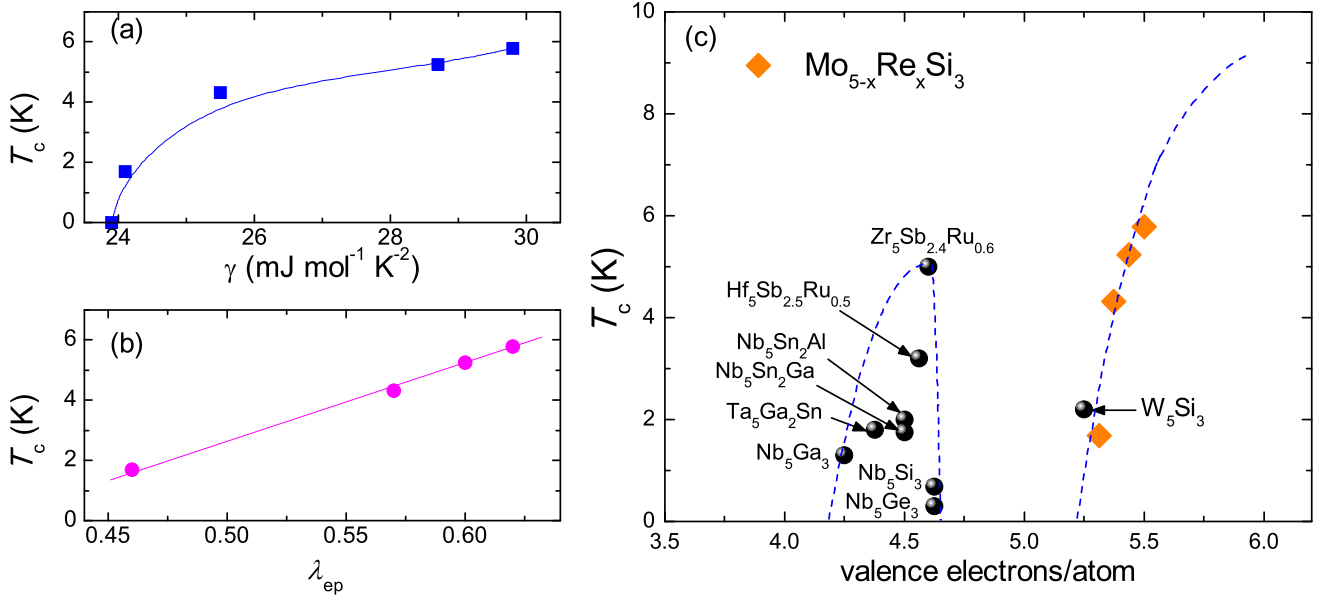}
\caption{
(a-b) $T_{\rm c}$ of the Mo$_{5-x}$Re$_{x}$Si$_{3}$ samples plotted as a function of $\gamma$ and $\lambda_{\rm ep}$, respectively.
(c) $T_{\rm c}$ of superconductors isostructural to Mo$_{5-x}$Re$_{x}$Si$_{3}$ plotted as a function of the average number of valence electrons per atom ratio. The two dashed lines are a guide to the eyes, showing that $T_{\rm c}$ of these superconductors follows the empirical Matthias rule.
}
\label{fig1}
\end{figure*}

Figure 2(a) shows the temperature dependence of resistivity ($\rho$) for the series of Mo$_{5-x}$Re$_{x}$Si$_{3}$ samples.
In all cases, $\rho$ decreases smoothly with decreasing temperature, indicative of a metallic behavior without any phase transition.
For undoped Mo$_{5}$Si$_{3}$ ($x$ = 0), $\rho$ remains finite down to 150 mK and the residual resistivity ratio (RRR) is about 3.
When increasing $x$ to 0.5, while the $\rho_{\rm 300K}$ value remains nearly unchanged, the RRR ratio is reduced to 1.5.
With further increasing $x$, the $\rho$($T$) curve shifts up almost rigidly, and the low temperature $\rho$ value for $x$ = 2 is nearly one order of magnitude larger than that for $x$ = 0.
This suggests that the incorporation of Re atoms leads to enhanced electron scattering.
Despite this enhancement, a clear drop to zero resistivity is observed for all Re-doped samples, and the resistive transition moves to higher temperatures as the increase of Re content.
From midpoints of the resistive transitions, the values of $T_{\rm c}$ are determined to be 1.67 K, 4.45 K, 5.29 K, and 5.78 K for $x$ = 0.5, 1, 1.5 and 2, respectively.
One may note that, in these samples, a higher $T_{\rm c}$ is correlated with a larger normal-state $\rho$.
However, it should be pointed that the resistivity of polycrystalline samples as used in the present study largely depends on the contribution of grain boundaries \cite{GB}.
Thus this correlation needs to be verified when single crystals become available.

The occurrence of superconductivity is corroborated by the magnetic susceptibility ($\chi$) and specific heat ($C_{\rm p}$) results shown in Figs. 2(c) and (d).
Large shielding fractions exceeding 100\% at 1.8 K (without correction for demagnetization effect due to the irregular sample shapes) and strong $C_{\rm p}$ anomalies are observed for the Mo$_{5-x}$Re$_{x}$Si$_{3}$ samples with $x$ $\geq$ 1, indicating bulk superconductivity. As for $x$ = 0.5, such anomaly is absent since it is expected below the lowest measurement temperature of 1.8 K.
The $T_{\rm c}$ values determined from the onset of diamagnetic $\chi$ are 4.63 K, 5.27 K and 5.78 K for $x$ = 1, 1.5 and 2, respectively.
On the other hand, entropy conserving constructions of the $C_{\rm p}$ anomalies yield $T_{\rm c}$ = 4.15 K, 5.11 K, 5.66 K, and $\Delta$$C_{\rm p}$/$\gamma$$T_{\rm c}$ = 1.42, 1.48, 1.64 for the same series of samples.
Note that the $T_{\rm c}$ values determined from transport and magnetic measurements are close to each other and  higher than that determined from the thermodynamic measurements.
The difference is less than 0.2 K for $x$ = 1.5 and 2 but attains a significantly larger value of $\sim$0.5 K for $x$ = 1, presumably due to lower sample homogeneity.

For all $x$ values, the normal-state $C_{\rm p}$ data are analyzed by the Debye model
\begin{equation}
C_{\rm p}/T = \gamma + \beta T^{2},
\end{equation}
where $\gamma$ and $\beta$ are the Sommerfield and phonon specific heat coefficients, respectively. Then the Debye temperature can be calculated as
\begin{equation}
\Theta_{\rm D} = (\frac{12\pi^{4}nR}{5\beta})^{1/3},
\end{equation}
where $n$ = 8 and $R$ = 8.314 J mol$^{-1}$ K$^{-1}$ is the molar gas constant.
Thus we obtain $\gamma$ = 24.2 mJ mol$^{-1}$ K$^{-2}$, 24.1 mJ mol$^{-1}$ K$^{-2}$, 25.5 mJ mol$^{-1}$ K$^{-2}$, 28.7 mJ mol$^{-1}$ K$^{-2}$, 29.8 mJ mol$^{-1}$ K$^{-2}$, and $\Theta_{\rm D}$ = 407 K, 406 K, 403 K, 397 K, 386 K for $x$ = 0, 0.5, 1, 1.5 and 2, respectively.
From $\Theta_{\rm D}$, the electron phonon coupling strength $\lambda_{\rm ep}$ can be estimated using the inverted McMillan formula \cite{McMillam},
\begin{equation}
\lambda_{\rm ep} = \frac{1.04 + \mu^{\ast} \rm ln(\Theta_{\rm D}/1.45\emph{T}_{\rm c})}{(1 - 0.62\mu^{\ast})\rm ln(\Theta_{\rm D}/1.45\emph{T}_{\rm c}) - 1.04},
\end{equation}
where $\mu^{\ast}$ is the Coulomb repulsion pseudopotential.
According to Ref. \cite{McMillam}, $\mu^{\ast}$ is usually between 0.1 and 0.13, the latter of which is taken for all transition metal elements.
Since the $N$($E_{\rm F}$) of Mo$_{5-x}$Re$_{x}$Si$_{3}$ are dominated by the contributions from Mo and Re atoms, $\mu^{\ast}$ value of 0.13 is reasonable.
Actually, we have checked the possible influence of the variation $\mu^{\ast}$ on the calculated results of $\lambda_{\rm ep}$ and found that it is insignificant.
Hence, for convenience, $\mu^{\ast}$ is assumed to be the same for all samples.
This gives $\lambda_{\rm ep}$ = 0.46, 0.57, 0.60, and 0.62 for $x$ = 0.5, 1, 1.5 and 2, respectively.

By subtracting the phonon contribution, we obtain the the normalized electronic specific heat $C_{\rm el}$/$\gamma$$T$ for $x$ $\geq$ 1 as shown in Fig. 2(e).
In order to fit the data, we employ a modified BCS model, or the so-called $\alpha$ model \cite{alphamodel}.
This model still assumes a fully gapped isotropic $s$-wave pairing while allows a variation of the coupling constant $\alpha$ $\equiv$ $\Delta_{\rm 0}$/$T_{\rm c}$, where $\Delta_{\rm 0}$ is the superconducting gap at 0 K.
Note that $\alpha_{\rm BCS}$ = 1.768 in the standard BCS theory \cite{BCSthoery}.
The $C_{\rm el}$/$\gamma$$T$ data for $x$ = 1.5 and 2 can be well reproduced by the model with $\alpha$ = $\alpha_{\rm BCS}$ and 1.88, respectively.
In comparison, the agreement with the data and the model with $\alpha$ = $\alpha_{\rm BCS}$ is slightly worse for $x$ = 1.
This is probably due to the sample inhomogeneity as noted above.
Nevertheless, it is emphasized that the overall quality of the fitting is well comparable to that reported in the literature \cite{HCfit}.
In addition, the $\alpha$ value for $x$ = 2 is larger than that for $x$ = 1 and 1.5, which is consistent with the trend of increasing $\lambda_{\rm ep}$ with Re content.
Taken together, these results suggest that the Mo$_{5-x}$Re$_{x}$Si$_{3}$ are weak coupling superconductors with a fully isotropic gap.
\begin{table*}
\caption{Measured chemical compositions, lattice and physical parameters of Mo$_{5-x}$Re$_{x}$Si$_{3}$.}
\renewcommand\arraystretch{1.3}
\begin{tabular}{p{3.7cm}<{\centering}p{2.4cm}<{\centering}p{2.4cm}<{\centering}p{2.4cm}<{\centering}p{2.4cm}<{\centering}p{2.4cm}<{\centering}p{0.5cm}}
\\
\hline % Top horizontal line
   Parameter  &  $x$ = 0  & $x$ = 0.5  & $x$ = 1 & $x$ = 1.5 & $x$ = 2 \\
\hline % In-table horizontal line
Measured Mo content & 5.06(8) & 	4.54(11)  	 & 4.01(11) & 3.65(6) & 3.17(13)	\\
Measured Re content & 0 & 	0.47(9)  	 & 1.09(3) & 1.43(3) & 1.83(6)	\\
Measured Si content & 2.93(8) & 	2.99(9)  	 & 2.90(12) & 2.91(7) & 2.99(11)	\\
$a$ ({\AA})& 9.648(1) & 	9.623(1)  	 & 9.590(1) & 9.569(1) & 9.545(1)	\\
$c$ ({\AA}) & 4.903(1) 	&  4.897(1)      & 4.911(1) & 4.914(1) & 4.911(1)   \\
$R_{\rm wp}$ & 5.7\% &	6.5\%  	 & 6.4\% & 7.1\% & 8.6\%	\\
$R_{\rm p}$ & 4.5\% &	5.0\% 	 & 4.8\% & 5.4\%& 5.9\%	\\
GOF & 1.33&	1.51 	 & 1.43& 1.58& 1.77	\\
$T_{\rm c}$($\rho$) (K) &$-$  & 1.67   	 & 4.45 & 5.29 & 5.78 	\\
$T_{\rm c}$($\chi$) (K) &$-$  & $-$   	 & 4.63 & 5.27 & 5.78 	\\
$T_{\rm c}$($C_{\rm p}$) (K) &$-$  & $-$   	 & 4.15 & 5.11 & 5.66 	\\
$\Delta$$C_{\rm p}$/$\gamma$$T_{\rm c}$ &$-$& 	$-$  & 1.42 & 1.48 & 1.64	\\
$\gamma$ (mJ molatom$^{-1}$ K$^{-2}$) & 23.9& 	24.1  & 25.5 & 28.7 & 29.8	\\
$\Theta_{\rm D}$ (K) & 407 & 	406  	 & 403 & 397 & 386	\\
$\lambda_{\rm ep}$ & $-$ & 	0.46  	 & 0.57 & 0.60 & 0.62	\\
$B_{\rm c2}$(0) (T)& $-$ & $-$  	 & 4.2 & 6.5 & 8.5	\\
$\xi_{\rm GL}$ (nm)& $-$ & 	$-$  	 & 8.7 & 7.1 & 6.2	\\
\hline% In-table horizontal line
\hline % Bottom horizontal line
\end{tabular}
\label{Table3}
\end{table*}

\begin{figure*}
\includegraphics*[width=17cm]{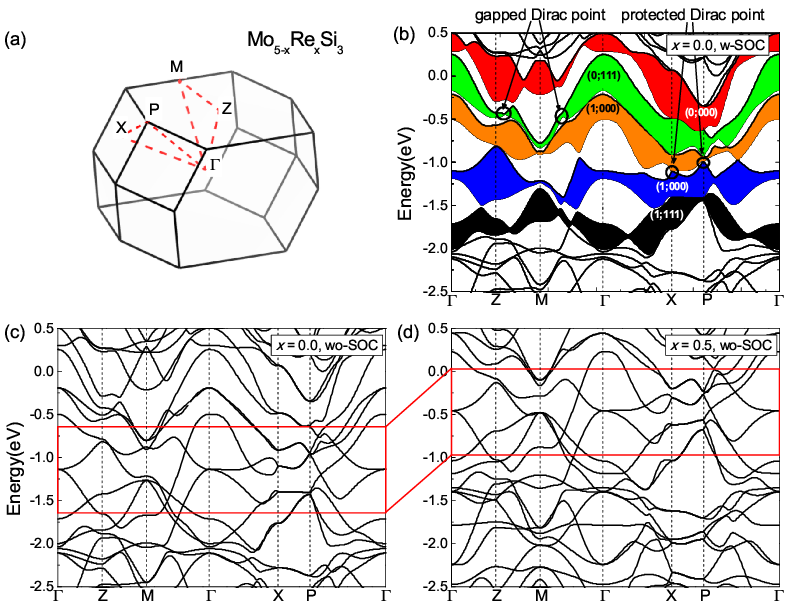}
\caption{
(a) Brillouin zone of Mo$_{5-x}$Re$_{x}$Si$_{3}$ with high symmetry points and lines. (b) Bulk band structure of Mo$_{5}$Si$_{3}$ ($x$ = 0) with SOC included. Gapped and protected Dirac points are highlighted. Colorful regions indicate where gap opens due to SOC, which allows us to calculate the $Z_{2}$ invariants. Taking the green one as an example, if the top band is unoccupied while the bottom band is occupied, we find a weak TI phase with $Z_{2}$ invariants being (0;111). (c, d) Band structures of $x$ = 0 and 0.5, respectively, without SOC included. A comparison of the bands in the energy regions indicated by the red frames shows that they look very similar to each other.
}
\label{fig1}
\end{figure*}

Figures 3(a)-(c) show the temperature dependence of resistivity under various magnetic fields for the Mo$_{5-x}$Re$_{x}$Si$_{3}$ samples with $x$ = 1, 1.5 and 2, respectively.
Increasing the field leads to a gradual suppression of the resistive transition, and the same 50\%$\rho_{\rm N}$ criterion is used to determine $T_{\rm c}$ under field.
The resulting upper critical field ($B_{\rm c2}$)-temperature phase diagrams are summarized in Fig. 3(d).
As can be seen, the $B_{\rm c2}$($T$) data for $x$ = 1 are well fitted by the Werthamer-Helfand-Hohenberg (WHH) model \cite{WHH}, yielding a zero-temperature upper critical field $B_{\rm c2}$(0) = 4.4 T.
In contrast, the data for the other two samples show upward deviation from the WHH behavior especially at low temperature.
As such, the $B_{\rm c2}$(0) values are estimated by extrapolating with the Ginzburg-Landau (GL) model
\begin{equation}
B_{\rm c2}(T) = B_{\rm c2}(0)\frac{1-t^{2}}{1+t^{2}},
\end{equation}
where $t$ = $T$/$T_{\rm c}$.
This gives $B_{\rm c2}$(0) = 6.5 T and 8.5 T for $x$ = 1.5 and 2, respectively.
With these $B_{\rm c2}$(0) values, the GL coherence length $\xi_{\rm GL}$ can be calculated by the equation
\begin{equation}
\xi_{\rm GL} = \sqrt{\frac{\Phi_{0}}{2\pi B_{\rm c2}(0)}},
\end{equation}
where $\Phi_{0}$ = 2.07 $\times$ 10$^{-15}$ Wb is the flux quantum.
Hence $\xi_{\rm GL}$ values of 8.7 nm, 7.1 nm and 6.2 nm are obtained for $x$ = 1, 1.5 and 2, respectively.

Now we examine the correlation between $T_{\rm c}$ and the physical parameters $\gamma$ and $\lambda_{\rm ep}$ to gain some insight into the pairing mechanism of Mo$_{5-x}$Re$_{x}$Si$_{3}$.
As can be seen from Figs. 4(a) and (b), $T_{\rm c}$ tends to increase with the increases of both $\gamma$ and $\lambda_{\rm ep}$, which is consistent with the BCS theory \cite{BCSthoery}. However, the $\gamma$ value is nearly the same for $x$ = 0 and 0.5, and hence the emergence of superconductivity is unlikely due to a change in $N$($E_{\rm F}$). Indeed, when increasing $x$ from 0.5 to 2, $T_{\rm c}$ is enhanced by more than a factor of 3, while $\gamma$ is increased by only $\sim$24 \%.
Remarkably, $T_{\rm c}$ of Mo$_{5-x}$Re$_{x}$Si$_{3}$ exhibits a linear dependence on $\lambda_{\rm ep}$, suggests that it is primarily governed by the electron-phonon coupling strength.
Previous study shows that superconductivity emerges in the isostructural compound Mo$_{5}$Ge$_{3}$ after neutron irradiation, which dose not modify its $\gamma$ but results in phonon softening \cite{Mo5Ge3}. According to McMillan \cite{McMillam},
\begin{equation}
\lambda_{\rm ep} \equiv \frac{N(0)\langle I^{2}\rangle}{M\langle \omega^{2}\rangle},
\end{equation}
where $N$(0) is the bare density of states at the Fermi level, $M$ is the atomic mass, $\langle$$I^{2}$$\rangle$ and $\langle$$\omega^{2}$$\rangle$ are the averaged electron-phonon matrix elements and phonon frequencies, respectively. For irradiated Mo$_{5}$Ge$_{3}$, $\lambda_{\rm ep}$ is enhanced mainly due to the decrease of $\langle$$\omega^{2}$$\rangle$ since $\Theta_{\rm D}$ reduces significantly from 377 K to 320 K after irradiation \cite{Mo5Ge3}. In contrast, the values of  $N$(0), $M$ and $\langle$$\omega^{2}$$\rangle$ are nearly the same for the Mo$_{5-x}$Re$_{x}$Si$_{3}$ samples with $x$ = 0 and 0.5. Thus the increase in $\lambda_{\rm ep}$ in this case can only be due to the enhancement of electron-phonon matrix elements. It is thus clear that the effect of Re doping is different from that of irradiation in inducing superconductivity.
In this respect, the phonon spectrum as well as its evolution with Re doping in Mo$_{5-x}$Re$_{x}$Si$_{3}$ is definitely of interest for further investigations.

Figure 4(c) shows the $T_{\rm c}$ dependence on $e$/$a$ for Mo$_{5-x}$Re$_{x}$Si$_{3}$ together with previously known isostructural superconductors \cite{Nb5Sn2Al}.
According to the Matthias rule, $T_{\rm c}$ of intermetallic superconductors is expected to exhibit two maxima at $e$/$a$ close to 4.7 and 6.5 \cite{MTrule}.
Indeed, the $e$/$a$ values for most of the superconductors isostructural to Mo$_{5-x}$Re$_{x}$Si$_{3}$ are in the range of 4.25-4.625 with a maximum $T_{\rm c}$ of 5 K observed at $e$/$a$ = 4.6.
As for Mo$_{5-x}$Re$_{x}$Si$_{3}$, the $e$/$a$ values of 5.312-5.5 fall significantly above this range and $T_{c}$ shows a rapid increase with increasing $e$/$a$, which is consistent with the presence of another $T_{\rm c}$ maximum at higher $e$/$a$. In this regard, a higher $T_{\rm c}$ might be expected by substituting Mo in Mo$_{5}$Si$_{3}$ with elements that can supply more valence electrons than Re, such as Ru, Rh, Pd, Os, Ir, and Pt. On the other hand, this can serve as a useful guide for the search of superconductivity in related $A_{5}$$B_{3}$ compounds, such as Mo$_{5}$Ge$_{3}$ \cite{Mo5Ge3}, Cr$_{5}$Si$_{3}$ \cite{Cr5Si3}, Cr$_{5}$Ge$_{3}$ \cite{Cr5Ge3}, which may significantly enlarge this family of superconductors.

Since Mo$_{5}$Si$_{3}$ is predicted to be a topological semimetal, it is of natural interest to investigate the evolution of band topology with Re doping. Fig. 5(a) shows the Brillouin zone of Mo$_{5-x}$Re$_{x}$Si$_{3}$, and, as can be seen, there are five high symmetry points, $\Gamma$, Z, M, X and P. The calculated band structure with SOC included for Mo$_{5}$Si$_{3}$ ($x$ = 0) is depicted in Fig. 5(b). The result indicates that several bands are crossing the Fermi level ($E_{\rm F}$), consistent with the metallic nature. Without considering SOC, there are two nodal lines: one is surrounding M point at Z-M-$\Gamma$ plane and another one is the X-P high symmetry line. When turning on SOC, two gapped Dirac points are located at $\sim$0.45 eV below the $E_{\rm F}$ and the two protected Dirac points are located exactly a the X and P points. This opens a gap between the valence and conduction bands (see the green region). Therefore, we can still calculate the four $Z_{2}$ invariants ($\nu_{0}$;$\nu_{1}$, $\nu_{2}$, $\nu_{3}$) to be (0;111), indicating that the undoped system is in the weak topological insulator (TI) phase. Note that the fourfold degeneracy of bands at X/P point is protected by various symmetries (mirror $M_{z}$, inversion and time-reversal symmetries) even under SOC. This, together with the nontrivial topological index, demonstrates that Mo$_{5}$Si$_{3}$ is a Dirac nodal line semimetal, confirming the previous studies \cite{Mo5Si3prediction1,Mo5Si3prediction2,Mo5Si3prediction3}.

As indicated by above XRD results, the Re dopants are distributed homogeneously in the lattice and hence the point group symmetries remain unchanged. In view of this, virtual crystal approximation have been adopted to elucidate the band structure variation induced by Re doping.
Figs. 5(c) and (d) show the band structures without considering SOC for $x$ = 0 and 0.5, respectively. A comparison of these two plots uncovers that the bands near $E_{\rm F}$ for $x$ = 0.5 look very similar to those near $E$ $-$ $E_{\rm F}$ = $-$0.6 eV for $x$ = 0. This indicates that the band character and band ordering of Mo$_{5-x}$Re$_{x}$Si$_{3}$ with $x$ $>$ 0 resemble closely the valence bands of undoped Mo$_{5}$Si$_{3}$ so that $Z_{2}$ invariants of the former can be facilely checked. As also seen in Fig. 5(b), the $Z_{2}$ invariants are (1;000) when $E$ $-$ $E_{\rm F}$ ranges from $-$0.25 eV to $-$1.5 eV and (1;111) when $E$ $-$ $E_{\rm F}$ ranges from $-$1.5 eV to $-$2.0 eV. Clearly, this indicates that the Re-doped Mo$_{5}$Si$_{3}$ are mostly in the strong TI phase, though the band topology changes with the Re doping level. In passing, we have also calculated the density of states as a function of energy for Mo$_{5-x}$Re$_{x}$Si$_{3}$ (see Fig. S2 of the Supporting Information). The evolution of calculated $N$($E_{\rm F}$) with Re content is consistent with that from the $C_{\rm p}$ results.

From above results, one can see that doped silicides Mo$_{5-x}$Re$_{x}$Si$_{3}$ with 0.5 $\leq$ $x$ $\leq$ 2 not only exhibit fully gapped superconductivity, but also possess non-trivial band topology characterized by strong $Z_{2}$ invariants. These together render them promising candidates for TSC \cite{TSCreview1,TSCreview2,TSCreview3}. In this regard, future angle resolved photoemission spectroscopy and scanning tunneling microscopy studies are called for to probe the band structure and superconducting gap directly. In particular, it is of significant interest to see whether zero-bias conductance peaks can be found within the magnetic vortex cores for certain $x$ values, which is a hallmark of the existence of Majorana bound states \cite{ZBCP1,ZBCP2,ZBCP3}. Also, thin film growth of these silicides is worth pursuing.
On one hand, by choosing proper substrates, a compressive or tensile strain could be generated, which may increase $\lambda_{\rm ep}$ and hence enhance $T_{\rm c}$.
On the other hand, electrostatic gating of these films can be used to tune the $E_{\rm F}$ continuously.
This would allow us to control both $T_{\rm c}$ and band topology, and may find applications in Si-based superconducting devices.

\section{IV. Conclusions}
In summary, we have discovered Re-doping induced superconductivity in the topological semimetal Mo$_{5}$Si$_{3}$. The substitution of Re for Mo results in a shrinkage of the Mo$_{5-x}$Re$_{x}$Si$_{3}$ unit cell along the $a$-axis up to the solubility limit of $x$ $\approx$ 2. Meanwhile, fully gapped superconductivity is observed for 0.5 $\leq$ $x$ $\leq$ 2 and $T_{\rm c}$ increases monotonically with the Re content. At $x$ = 2, $T_{\rm c}$ reaches a maximum of 5.87 K, which is the highest among superconductors of this structural type.
Our analysis shows that the emergence of superconductivity as well as the trend in $T_{\rm c}$ is mainly attributed to the enhancement of electron-phonon coupling.
In addition,  $T_{\rm c}$ is found to increase as the increase of $e$/$a$, implying that a higher $T_{\rm c}$ may be achieved by adding more valence electrons.
First-principles calculations show that Re doping turns the system into strong TI phase, which is characterized by $Z_{2}$ invariants (1;000) or (1;111) depending on the doping level.
The combination of a fully gapped superconducting state and nontrivial band topology makes Mo$_{5-x}$Re$_{x}$Si$_{3}$ a promising candidate for TSC.
Our study suggests that transition metal silicides may offer a new playground to explore topological superconductivity, which might facilitate the creation and control of Majorana fermions in silicon-based devices.

\section*{ACKNOWLEGEMENT}
We acknowledge financial support by the foundation of Westlake University.
The work at Zhejiang University is supported by National Key Research Development Program of China (No.2017YFA0303002), National Natural Science Foundation of China (No.11974307, No.61574123), Zhejiang Provincial Natural Science Foundation (D19A040001), Research Funds for the Central Universities and the 2DMOST, Shenzhen Univeristy (Grant No.2018028).


\begin{thebibliography}{00}
\bibitem{orderedmodel}
Note that we have also tried structural refinement based on an ordered structural model.
For example in Mo$_{4}$ReSi$_{3}$ ($x$ = 1), the Re atoms are set to occupy the Mo(2) position.
Nevertheless, the resulting $R_{\rm wp}$ and $R_{\rm p}$ factors are larger than those for the model with a disordered distribution of the Re atoms.
\end{thebibliography}
\end{document}